\documentclass[aps,twocolumn,epsfig,showpacs]{revtex4}
\usepackage{graphicx}
\usepackage{epsfig}
\begin{document}
\title{Effect of shear force on the separation of double stranded DNA}
\author{Rakesh Kumar Mishra, Garima Mishra, M. S. Li$^{1}$ and 
Sanjay Kumar }
\affiliation{Department of Physics, Banaras Hindu University,
     Varanasi 221 005, India \\
$^{1}$~Institute of Physics, Polish Academy of Sciences, Al. Lotnikow
32/46, 02-668 Warsaw, Poland }
%\date{\today}
\begin{abstract}
Using the Langevin Dynamics simulation, we have studied the effects
of the shear force on the rupture of short double stranded DNA at 
different temperatures. We show that the rupture force increases linearly with
the chain length and approaches to the asymptotic value in 
accordance with the  experiment. The qualitative nature of these
curves almost remains same for different temperatures but with a shift 
in the force. We observe three different regimes in the extension of 
covalent bonds (back bone) under the shear force.

\end{abstract}
\maketitle

%\section{introduction}
Inter- and intra- molecular forces are key to the stability of DNA and 
biological processes {\it e.g.} transcription, replication, slippage 
etc. \cite{albert,israel}. Up to now, understanding of these forces was 
possible through the indirect physical and thermodynamical measurements like 
crystallography, light scattering, nuclear magnetic resonance spectroscopy etc. 
\cite{Wartel_Phys.Rep85}.
Single molecule force spectroscopy (SMFS) experiments have directly measured
these forces and provided unexpected insights into the strength of the forces 
driving these biological processes as well as determined various  
interactions responsible for the mechanical stability of DNA structures
\cite{Smith_Science92,cluzel,Lee_Science94,kumarphys}. 
%Given the experimental setup of SMFS, it was thought that the interactions 
%detected would be mostly of a mechanical nature and can be computed by 
%knowing the magnitude of the applied force. 
With the increasing number of experiments and insights gathered so far, it 
has become clear that the measurement of molecular interactions not only depends 
on the magnitude of the applied force, but also depends how and where the 
force was applied \cite{kumarphys,Bockelmann,Bock,Strunge,Irina,
prentiss1,hatch,Cludia,gaub}.

A major concern is now to understand whether all these interactions contribute
at the same moment or they have different life times. In order to understand
this, a force has been applied perpendicular to the helix direction (DNA 
unzipping) and along the helix direction (rupture and slippage) as shown
in Fig. 1 \cite{kumarphys,Bockelmann, Bock,Strunge,Irina, prentiss1,hatch,Cludia,
gaub,cocco}. In case of unzipping of double stranded DNA (dsDNA), the critical
force is found to be independent of the length of DNA and the loading rate 
\cite{Bockelmann, Bock}. This may be understood theoretically that at the 
centre point of fork (Fig. 1b), the applied force only breaks a base pair at a time 
and hence it remains independent of the loading rate and length. However, when 
a force (up to 65 pN)
is applied along the helix direction (shear force), the length of the dsDNA
increases and the force-extension ($f-x$) curve can be described 
by the worm like chain (WLC) model \cite{wlc}. In the high force regime ($ >65$ 
pN), the dsDNA can be overstretched about 1.7 times of the B-form contour 
length and a phase transition occurs from the B-form to 
a stretched or S-form \cite{Rief,smith,Morfill}. Recently, van Mameren {\it et al}. 
have studied DNA stretching with or without DNA binding ligands and 
demonstrated that overstretching comprises a gradual conversion from dsDNA 
to ssDNA and it should be interpreted in terms of force induced DNA 
melting  \cite{mameren}. 

\begin{figure}[t]
%\centerline{\epsfig{file=forcedirection.eps.eps, scale=0.4}}
\includegraphics[width=2.in]{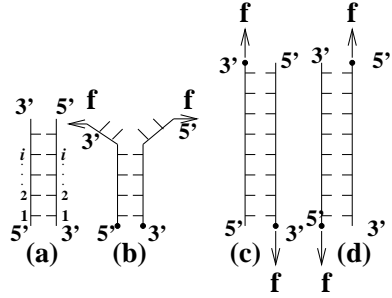}
\caption{Schematic representation of dsDNA: (a) dsDNA in zipped form; 
(b) Unzipping of dsDNA by the force ($f$) applied at one end $(5'-3')$; 
(c and d) Shear force along the chain applied at the opposite ends 
($3'-3'$ or $5'-5'$) of the dsDNA.}
\label{fig-1}
\end{figure}

For a short dsDNA, if the applied shear force increases, the dsDNA separates 
into two single strands at some critical force. This phenomenon has been 
identified as rupture \cite{Lee_Science94, Strunge}. The unbinding force strongly depends on the 
pulling end and is much larger than the unzipping force \cite{Cludia,Lee_Science94,Strunge,lavery}. 
Neher and Gerland studied the dynamics of dissociation of the two strands and 
found the expression for the critical force \cite{nehar}. Expressing the bond energy and 
the base pairing energy in the form of harmonic oscillators in the ladder model of dsDNA of
length $L$, de Gennes \cite{degennes} proposed the maximum force required for the rupture 
\begin{equation}
f_c= 2 f_1 (\chi^{-1} \tanh(\chi \frac{L}{2})+1).
\end{equation}
Here, $f_1 $ is the force required to separate a single base pair, which
is same for the homo-sequence and ${\chi^{-1} = \sqrt{Q/2R}}$
 is the de Gennes characteristic length over which differential force is distributed.
Here, Q and R are the spring constants, characteristic of stretching of backbone and 
hydrogen bonds, respectively.

Recently, Danilowicz {\it et al.} \cite{hatch} 
systematically studied the DNA rupture by varying the length of 
dsDNA. The critical shear force is found to increase linearly 
up to a certain length and approaches the asymptotic value ($\approx 62$ pN),
which is in good agreement with the de Gennes prediction. 
It was argued that the covalent bonds (backbone) and 
the hydrogen bonds involved in the base pairing will be stretched under the applied 
force. The differential force will approach to zero at the length 
$\chi^{-1}$, if one moves in from the either side. 
However, no experimental effort has been made to study the effect of shearing force 
on the stretching of covalent bonds and hydrogen bonds in side the characteristic length 
$\chi^{-1}$.  Moreover, in the description of de Gennes model \cite{degennes} or 
subsequently improved model by Chakrabarti and Nelson \cite{nelson}, effect of thermal  
 fluctuation has been ignored, whereas all the rupture experiments were generally performed at finite 
temperature. The aim of this manuscript is to study the effect of temperature ($T$) on the 
rupture and consequences of differential force on the distribution of 
extension in bond lengths and hydrogen bonds near the rupture.

We use Langevin  Dynamics (LD)simulation to investigate mechanical and 
physical properties related to the rupture of DNA 
\cite{Allen,Smith,Kouza,MSLi_BJ07}. Since, rupture time is 
of the order of milliseconds to seconds, 
an atomistic simulation of longer chain in the solvent is computationally 
difficult \cite{netz,pm}. We have used a coarse-grained model 
\cite{kumarphys,Kouza,MSLi_BJ07,janke} of the flexible polymer chain 
to model a DNA, which allows us to study a larger system size and events of 
a longer time scale.  A chain in the model consists of bead units connected by 
effective bonds characterized by the stiff springs. Each effective bond 
represents several chemical bonds ({\it e.g.} sugar phosphate etc. ) along 
the chain backbone. 
%A Lennard-Jones (LJ) potential is used for the hydrogen bonding
%that forms base pairs (bps) between complimentary nucleotides. 
The energy of the model system is given  by 
%\cite{Li,Kouza,MSLi_BJ07}
%\begin{widetext}
\begin{eqnarray}
%E = \sum_{i=1}^{N-1}k(d_{i,i+1}-d_0)^2+\sum_{i=1}^{N-2}\sum_{j>i+1}^{N}4(\frac{C}{d_{i,j}^{12}}-\frac{A_{ij}}{d_{i,j}^6}),
& & E   =  {\sum_{l=1}^2\sum_{j=1}^N}k(u_{j+1,j}^{(l)}-d_0)^2
+{\sum_{l=1}^2\sum_{i=1}^{N-2}\sum_{j>i+1}^N}4\left(\frac{C}{{u_{i,j}^{(l)}}^{12}}\right) \nonumber \\
 & & + {\sum_{i=1}^N\sum_{j=1}^N}4\left(\frac{C}{(|\vec u_i^{(1)}-\vec u_j^{(2)})|^{12}}-
\frac{A}{(|\vec u_i^{(1)}-\vec u_j^{(2)}|)^6}\delta_{ij}\right),
%+{\sum_{l=1}^2\sum_{i=1}^{N-2}\sum_{j>i+1}^N}4(\frac{C}{{u_{i,j}^{(l)}}^{12}})
\end{eqnarray}
%\end{widetext}
where $N$ is the number of beads in each strand. $\vec u_i^{(l)}$ represents the position of $i^{th}$ bead 
on $l^{th}$ strand. In present case, $l=1(2)$ corresponds to first (complimentary) 
strand of  dsDNA. 
The distance between intra strand beads, $u_{i,j}^{(l)}$, is defined as
$|\vec u_i^{(l)}-\vec u_j^{(l)}|$.
The harmonic (first) term with spring constant $k$ (=100) couples the adjacent beads
along the two strands. 
Second term takes care of excluded volume effect {\it i.e.} two beads
can not occupy the same space \cite{book}.
The third term, described by Lennard-Jones (LJ) potential, takes care of the 
mutual interaction between two strands.
The  first term of LJ potential (same as second term of Eq.2) will 
not allow the overlap of two strands. Here, we set $C = 1$ and $A=1$.
The second term of LJ potential corresponds to the base pairing between 
two strands. The base pairing interaction is restricted to the native contacts 
($\delta_{ij}=1$) only {\it i.e.}  $i^{th}$ base of $1^{st}$ strand forms pair with 
the $i^{th}$ base of $2^{nd}$ strand only as shown in Fig. 1a.  This is similar  to the 
Go model \cite{go}.
%By native we mean that ith base of one strand forms pair with the ith base of 
%2nd chain only as shown in Fig.1 a. 
 The parameter $d_0 (=1.12)$ corresponds to 
the equilibrium distance in the harmonic potential, which is close to the 
equilibrium position of the average LJ potential. In Eq. 2, 
we use dimensionless distances and energy parameters. The major advantage of 
this model is that the ground state energy of the system is known \cite{go}. 
%Therefore, equilibration is not an issue here, if
%one wants to study the dynamics under the applied force at low $T$.
The equation of motion is obtained from the following Langevin equation 
\cite{Allen,Smith,MSLi_BJ07} 
\begin{equation}
m\frac{d^2r}{dt^2} = -{\zeta}\frac{dr}{dt}+F_c+\Gamma,
\end{equation}
where $m$ and $\zeta$ are the mass of a bead and the friction 
coefficient, respectively. Here, $F_c$ is defined as $-\frac{dE}{dr}$ and 
the random force $\Gamma$ is a white noise \cite{Smith},
i.e., $<{\Gamma(t)\Gamma(t')}>=2\zeta T\delta(t-t')$. 
The choice of this dynamics keeps $T$ constant throughout the simulation 
for a given $f$. The equation of motion is integrated by using the $6^{th}$ order 
predictor-corrector algorithm with time step $\delta t$=0.025 \cite{Smith}.  
The results are averaged over many trajectories. The equilibration 
has been checked by monitoring the stability of data against at least ten 
times longer run. We have used $2\times10^9$ time steps out of which 
first $ 5\times 10^8$ steps have not been taken in the averaging.

\begin{figure}[t]
\includegraphics[width=3.4in]{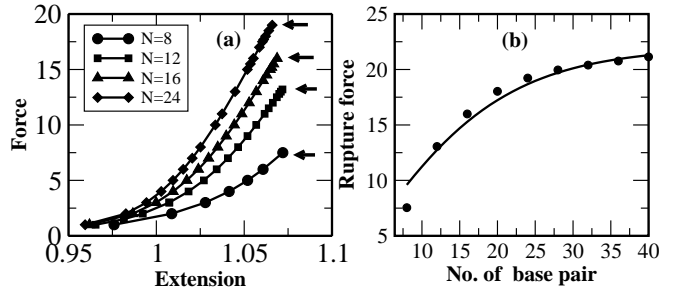}
\caption{(a) Force {\it vs} extension curves for different chain lengths
% in the constant force ensemble. 
Arrows indicate the maximum force, where the number of 
contacts approaches to zero.
% and two chains are separated. 
For the sake of comparison, we have normalized the extension 
by its contour length (at f = 0).
(b) Figure shows the variation of rupture force with length. The solid
line corresponds to a fit of Eq. 1.
% with $f_1 = 1$ and $\chi =0.1$.
Solid circles represent the value obtained through the simulation. 
%A nice agreement with theory \cite{degennes} is apparent from the plot.
}
\label{fig-2}
\vspace {0.5cm}
\end{figure}
%\section{Constant force simulation}

In the constant force ensemble, we add an energy $-\vec{f}.\vec{x}$ to the 
total energy
of the system given by Eq. 2. We calculate the reaction coordinate 
$x$ (extension)  for different values of $f$. The $f-x$ curves 
(Fig. 2a) show the entropic response at low forces
and remain qualitatively similar to the one seen in experiments 
\cite{Lee_Science94,Strunge,Irina,gaub}. 
We identify 
the rupture force as a maximum force, where the number of intact base pairs  
 suddenly goes to zero.  In Fig. 2b, we show the 
rupture force as a function of the chain length at low temperature. 
It is evident from this 
plot that the rupture force approaches to an asymptotic value for the chain
length greater than 20, which is in accordance with the experiment \cite{hatch}.

We expand the LJ potential given 
in Eq.2 around its equilibrium value. The coefficient of second 
(harmonic) term of its expansion corresponds to the elastic constant 
of the base-pairing. The de Gennes characteristic length \cite{degennes} 
for the present 
model is estimated to be $\approx 10$. Substituting the value of $f_1 (=1)$ and 
the above mentioned value of $\chi^{-1}$ in Eq. 1, we obtained the value of 
$f_c$ for a given length of dsDNA, which is shown by solid line in Fig 2b.
One can notice a nice agreement between the simulation and the value predicted 
by Eq. 1 \cite{degennes}.

\begin{figure}[t]
%\centerline{\epsfig{file=forcedirection.eps.eps, scale=0.6}}
%\includegraphics[width=2.4in]{Fig4.eps} \\
%\vspace{0.1in}
%\includegraphics[width=2.4in]{Fig4b.eps}
\includegraphics[width=3.4in]{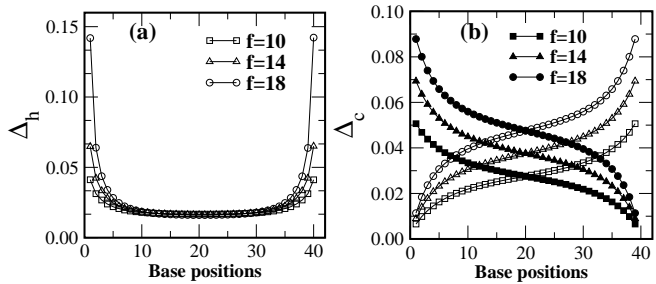}
\caption{(a) Variation of extension in hydrogen bond length ($\Delta_h$) 
along the chain at three different forces. 
%one can notice that elongation in bond length at the ends
%are much larger than the middle, where the differential shearing force
%approaches to zero. 
(b) Figure shows the variation in extension of covalent 
bond length ($\Delta_c$) along
the chain. Here, open and solid symbols correspond to one 
strand and its complementary strand, respectively. 
%The curve has three regimes. 
%The larger extension in bond length correspond to the end
%where the  force is applied, whereas the minimum extension in bond length
%corresponds that the force is applied at the other end of the complimentary
%strands. The differential force here also approaches to zero, which
%can be seen from the middle of the chain.
}
\label{fig-3}
\end{figure}
%The distribution of extension in hydrogen bonds ($\Delta_h$) and  
%covalent bonds ($\Delta_c$) along the chain is difficult to obtain experimentally.
One of the important findings of the present simulation is the  distribution of 
stretching of hydrogen bonds ($\Delta_h$) and extension in the covalent bonds ($\Delta_c$)
for a wide range of force below the rupture, which are experimentally difficult to obtain. In Fig. 3a, 
we depict the variation of $\Delta_h$ with base position for the chain of length 40. 
%The characteristic de Gennes length for the present simulation is about 10 bases. 
From this plot, one can observe that the hydrogen bonds at extreme ends (up to 
$\approx 10$ bases) get stretched, whereas bases in the middle (above the de Gennes 
length $\approx 10\sim30$) remain same indicating that the differential shear 
force approaches to zero in this region. In Fig. 3b, we show the variation of $\Delta_c$ 
with the base position.  All curves have three distinctively different regions. 
One can observe that bonds near the pulling end (say 5'-end) get stretched more and 
decreases gradually. However, 
when one approaches the other end (i.e. 3'-end), there is a change in the 
slope and the 
extension is quite less compare to the middle one. It should be noted  
that 3'-end is near to 5'-end of the other chain, where a similar force is
also applied. Since, dsDNA is in the zipped state, the applied force 
at 5'-end
of one strand also pulls the other strand along the opposite direction, 
which causes a relatively slower increase.

\begin{figure}[t]
\includegraphics[width=3.4in]{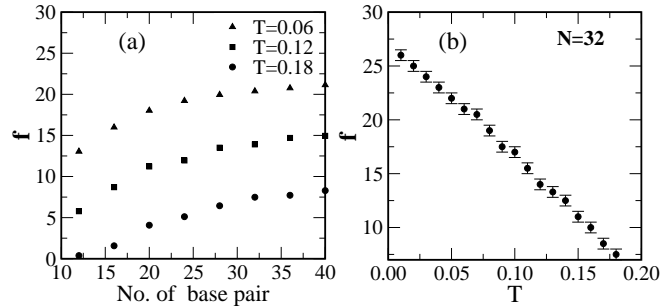}
\caption{(a) Variation of the rupture force with chain length at different T. 
(b) Figure shows the DNA rupture force-temperature diagram for chain 
length 32.
}
\label{fig-4}
\end{figure}
In model studies either temperature is set to zero \cite{degennes} or thermal 
 fluctuation is ignored \cite{nelson} as a result,
 the rupture force defined in Eq. 1 is independent of temperature. 
However, all rupture experiments usually performed at room temperature
\cite{Strunge,hatch}, therefore, it is desirable to understand role of entropy, which 
may play a significant role at higher $T$. The thermodynamics of force 
induced melting can be obtained from the following relation \cite{rouzina,singh}:
\begin{equation}
-fx= \Delta H - T\Delta S,
\end{equation}
where $H$ is the enthalpy and S is entropy of the ruptured chains. Setting 
$x$ equal to unity and replacing the value of $\Delta H$ by the  value of 
the rupture force at $ T= 0$, Eq. 4 can be written as
\begin{equation}
-f = f_c - T\Delta S,
\end{equation}
where $f_c$ is given by Eq. 1. In the thermodynamic limit, value of $S$ may be
analytically estimated \cite{book} or can be obtained from the experiments \cite{santalucia}.
However, for the finite chain length, one has to resort on numerical 
techniques to get the value of $S$.  
It is possible to study the effect of temperature on the rupture force 
in the present setup. In Fig. 4a, we show the dependence of the rupture force on the
length at different temperatures. The qualitative nature of the curves 
obtained at different $T$ remain similar to de Gennes plot (Fig. 2b),
with a  shift showing that the rupture force decreases with $T$. From Fig. 4a, 
one can notice that for chain length 32, the rupture force has approached  
its  asymptotic value for all $T$. In Fig. 4b, we depict the force-temperature 
diagram for the DNA rupture for the chain length 32. A linear dependence on 
temperature can be noticed, which is in accordance with Eq. 5. 
Usually rupture experiments are performed well below  the melting
temperature, which correspond that the DNA is in the zipped state.
Therefore, in Fig. 4b, maximum temperature has been set slightly below the 
melting temperature i.e. $T_m =0.23$ at $f = 0$.
%\section{Conclusions}

In this paper, we have studied the effect of shear force and temperature
on the rupture of dsDNA. We show that  the shear force increases 
linearly with the length of the DNA and then approaches to the asymptotic 
value. In the lattice model, the bond length is constant (stiff), 
therefore, covalent bonds and hydrogen bonds will not be stretched. As
a result, the rupture force increases linearly with length \cite{singh} 
in the lattice model or in the models, where bond length is considered 
as a constant.
Our simulations confirm that, one will not gain strength by increasing a 
larger pair sequence as predicted by de Gennes \cite{degennes}. Interestingly, recent experiment also supports 
it \cite{hatch}.
The distribution of extension in
hydrogen bonds and covalent bonds for different forces
are shown in Fig. 3, where Fig. 3a clearly shows that the increase in the extension 
of the hydrogen bonds is limited up to the de Gennes length, which is
consistent with the strain profile obtained by Chakrabarti and Nelson
\cite{nelson}. Above this length, differential shear force approaches to 
zero, as a result there is no extension in the hydrogen bonds. 
We also find that the qualitative nature of dependence of rupture 
force on length remains same for different temperatures with a shift.
The rupture force decreases with temperature (Fig. 4b) as predicted 
by Eq. 5.

Our study shows that the de Gennes length remains independent of the 
applied force. The most surprising finding of the present simulation is 
revealed from Fig. 3b, which shows variation of extension in the 
covalent bonds along the chain for three different forces ($f=10,14,18$).
Although, in all these cases, the differential 
force approaches to zero and there is no relative increase in the bond 
length above the de Gennes length along the chain (Fig. 3b). However, unlike
the hydrogen bonds, there is a net increase in the extension in the covalent 
bonds, which depends linearly on the applied 
force. Nuclear magnetic resonance experiment \cite{nmr} or the atomistic 
simulation \cite{pm} should  be able to observe this.

It may be noted that the present simulation is carried out in the reduced unit.
It is possible to extract a rough estimate of the rupture force in the 
real unit. The free energy per base pair (including hydrogen bonding and base
stacking) of G-C is -1.4 kcal/mol, where stacking accounts probably half of 
this amount \cite{hydrogen_bond}. Since in A-T base pairing, only two hydrogen 
bonds are involved, one can take approximately two third value of the 
G-C free energy. For fixing the temperature scale, 
one can use DNA melting data where both stacking as well as hydrogen bonding 
are required. Whereas in rupture, we assume that there are breaking of hydrogen 
bonds only and hence stacking does not contribute significantly. In  
Ref. \cite{hatch}  hetero-sequence (50\% AT and 50\% GC) chain of different lengths
were considered. Therefore, we take approximately -0.6 kcal/mol per base of 
the zipped conformation and equate it with the complete unzipped state. The 
required force for the rupture is found to be approximately 3.5 pN per 
base pair, which is close to the one used in Ref. \cite{hatch}. So if one scales
the y-axis of Fig. 2b by 3.5 pN then our results are also in quantitative 
agreement with the  experiment \cite{hatch}.

%\section{acknowledgments}

We thank D. Giri for many helpful discussions on the subject. Financial 
supports from the DST, CSIR, India and the MSI, Poland (grant No 
202-204-234) are gratefully acknowledged.


\begin{thebibliography}{99}
\bibitem{albert} B. Alberts {\it et al.},
%, D. Bray, J. Lewis, M. Raff, K. Roberts and J. D. Watson, 
{\it Molecular Biology of the Cell}, (Garland Publishing: New York, 1994).
\bibitem{israel} J. N. Israelachvili, {\it Intermolecular and Surface Forces}
(Academic Press, London, 1992).
\bibitem{Wartel_Phys.Rep85} R. M. Wartell and A. S. Benight, Phys. Rep. {\bf 126}, 67 (1985).
\bibitem{Smith_Science92} S. B. Smith, L. Finzi and C. Bustamante, Science {\bf 258}, 1122 (1992).
\bibitem{cluzel} P. Cluzel {\it et al.}, Science {\bf 271}, 792 (1992).
\bibitem{Lee_Science94} G. U. Lee, L. A. Chrisey and R. J. Colton, Science {\bf 266}, 771 (1994).
\bibitem{kumarphys} S. Kumar and M. S. Li, Phys. Rep. {\bf 486}, 1 (2010).
\bibitem{Bockelmann} B. Essevaz-Roulet, U. Bockelmann and F. Heslot, PNAS {\bf 94}, 
11935 (1997).
\bibitem{Bock}  U. Bockelmann, B. Essevaz-Roulet and F. Heslot,  Phys. Rev. Lett. 
{\bf 79}, 4489 (1997).
\bibitem{Strunge} T. Strunz {\it et al.},
% K. Oroszlan, R. Sch\"{a}fer and H.J. G\"{u}entherodt,
PNAS {\bf 96}, 11277 (1999).
\bibitem{Irina} I. Schumakovitch {\it et al.}, Biophys. J. {\bf 82}, 517 (2002).
\bibitem{prentiss1} C. Danilowicz {\it et al.}, Phys. Rev. Lett. {\bf 93}, 078101
(2004).
\bibitem{hatch} K. Hatch {\it et al.}, Phys. Rev. E {\bf 78}, 011920 (2008).
\bibitem{Cludia} C. Danilowicz  {\it et al.},
%C. Limouse, K Hatch, A. Conover, V. W. Coljee, N. Kleckner, and M. Prentiss,
PNAS {\bf 106}, 13196 (2009).
\bibitem{gaub} F. K\"{u}eher {\it et al.}, 
%J. Morfill, R. A. Neher, K. Blank, and H. E. Gaub, 
Biophys. J. {\bf 92}, 2491 (2007).
\bibitem{cocco} S. Cocco,R.Monasson and JF.Marko, PNAS {\bf 98}, 8608 (2001).
\bibitem{wlc} J. F. Marko and E. D. Sigga, Macromolecules {\bf 28}, 8759 (1995).

\bibitem{Rief}M. Rief, H. C. Schaumann and H. E. Gaub, Nat. Struc. Bio. {\bf 6}, 346 (1999).
\bibitem{smith} S. B. Smith, Y. Cui and C. Bustamante, Science {\bf 271}, 795 (1996).
\bibitem{Morfill} J. Morfill {\it et al.}, Biophys. J.  {\bf 93}, 2400 (2007).
\bibitem{mameren} J. V. Mameren {\it et al.}, PNAS {\bf 106}, 18231 (2009).
\bibitem{lavery} A. Lebrun and R. Lavery, Nucleic Acids Res. {\bf 24}, 2260 (1996).
\bibitem{nehar} R. A. Nehar and U. Gerland,  Phys. Rev. Lett. {\bf 93}, 198102 (2004).
\bibitem{degennes} P. G. de Gennes, C. R. Acad. Sci. {\it -Series IV- Physics} {\bf 2}, 1505 (2001).
\bibitem{nelson} B. Chakrabarti and D. Nelson, J. Phys. Chem. B {\bf 113}, 3831 
(2009).
%\bibitem{nevo} R. Nevo {\it et al.} Embo Reports {\bf 6}, 482 (2005).
\bibitem{Allen}M. P. Allen and D. J. Tildesley, {\it Computer Simulations of Liquids}
(Oxford Science, Oxford, UK, 1987).
\bibitem{Smith} D. Frenkel and  B. Smit {\it Understanding Molecular Simulation} (Academic Press UK, 2002).
\bibitem{go}  N. Go and  H. Abe, Biopolymers {\bf 20}, 991 (1981); G. Mishra {\it 
et al.} J. Chem. Phys. {\bf 135}, 035102 (2011).
\bibitem{Kouza} M. Kouza {\it et al.}, Biophys. J. {\bf 89}, 3353 (2005).
\bibitem{MSLi_BJ07} M. S. Li, Biophys. J. {\bf 93}, 2644 (2007).
\bibitem{netz} T. Hugel {\it et al.},
%, M. Rief, M. Seitz, H. E. Gaub, R. R.  Netz, 
Phys. Rev. Lett. {\bf 94}, 48301 (2005)
\bibitem{pm} M. Santosh and P. K Maiti, J. Phys.: Cond. Matt. {\bf 21}, 034113 (2009).
\bibitem{janke} J. Schluttig, M. Bachmann and W. Janke, J. Comput. Chem. 
{\bf 29}, 2603 (2008).
\bibitem{book}P. G. de Gennes, {\it Scaling Concepts in Polymer Physics}
(Cornell Univ. Press, Ithaca, 1979).
\bibitem{rouzina} I. Rouzina and V. A. Bloomfield, Biophys. J. {\bf 80}, 882 (2001).
\bibitem{singh} A. R. Singh, D. Giri and S. Kumar, J. Chem. Phys. {\bf 132}, 235105(2010).
\bibitem{santalucia} J. Santalucia Jr., H. T. Allawi and P. A. Seneviratne,
Biochemistry {\bf 35}, 3555 (1996).
%\bibitem{bell} G.I. Bells, Science {\bf 200}, 618 (1978).
%\bibitem{evan} E. Evans and  K. Ritchie, Biophys. J. {\bf 72}, 1541 (1997).
%\bibitem{kramer} H.A. Kramers, Physica (Utrecht) {\bf } 7 284 (1940).
\bibitem{nmr} W. Saenger, {\it Principles of Nucleic Acid Structure}
(Springer-Verlag, Berlin, 1984).
\bibitem{hydrogen_bond} E. T. Kool, Annu. Rev. Biophys. Biomol. Struct. 
{\bf 30}, 1 (2001).

\end{thebibliography}
\end{document}